# Study of Mechanical Response in Embossing of Ceramic Green Substrate by Micro-Indentation

Y. C. Liu and Xuechuan Shan[+]

Singapore Institute of Manufacturing Technology (SIMTech),
71 Nanyang Drive, Singapore 638075

[+] Corresponding Email: xcshan@simtech.a-star.edu.sg (Xuechuan Shan)

*Abstract*- **Micro-indentation test with a micro flat-end cone indenter was employed to simulate micro embossing process and investigate the thermo-mechanical response of ceramic green substrates. The laminated low temperature co-fired ceramic green tapes were used as the testing material; the correlations of indentation depth versus applied force and applied stress at the temperatures of 25 °C and 75°C were studied. The results showed that permanent indentation cavities could be formed at temperatures ranging from 25 °C to 75 °C, and the depth of cavities created was applied force, temperature and dwell time dependent. Creep occurred and made a larger contribution to the plastic deformation at elevated temperatures and high peak loads. There was instantaneous recovery during the unloading and retarded recovery in the first day after indentation. There was no significant pile-up due to material flow observed under compression at the temperature up to 75 °C. The plastic deformation was the main cause for formation of cavity on the ceramic green substrate under compression. The results can be used as a guideline for embossing ceramic green substrates.**

## I. INTRODUCTION

Low temperature co-fired ceramic (LTCC) has been used for many years in the microelectronics packaging industry. One of the important advantages of LTCC technology is that it can be used to generate 3D structures using multi-layers of tapes and it is a way to create multi-layer circuits with the help of single tapes and applying conductive, dielectric and resistive materials. It can be used for various applications such as telecommunication devices, microwave components, sensors, actuators, microfluidics, meso-system module, bio-medical devices, etc [1]. The green ceramic tapes are glass-ceramic composite materials. They include ceramic filler that is usually alumina, glass frit binder and organic vehicle for binding and viscosity control of the tape before sintering. Green ceramic tapes can be manipulated for fabricating 3D structures. The traditional techniques for structuring green ceramic tapes are micro punching, milling and laser machining, etc [2, 3].

Recently micro embossing technique has been applied to create 3D structures in ceramic multi-layers [1, 4, 5]. This opened a new possibility of patterning features with micrometer scales on green tapes. Micro embossing or imprinting is usually used for patterning microstructures on a polymeric substrate [6-8]. It uses a master mold to press onto a substrate material under a certain pressure and a temperature, which is usually above the glass transition temperature of the material, to transfer the mold patterns onto the substrate. After the temperature returns below its glass transition temperature, the mold is separated from the substrate, and the substrate material retains the shape of the mold patterns. In this process, temperature plays an important role to both the local-area fidelity and global uniformity of the structure formation. Higher embossing temperature could improve the fidelity of the structures but could cause serious thermal stress during demolding, which results in poor global flatness after demolding. With lower embossing temperature, the global flatness could be improved significantly [9]. Hence, a compromise between local fidelity of the embossed patterns and global flatness of the substrate has to be considered. Therefore, study of micro embossing at low temperature for green ceramic tapes is necessary.

The recent development in micro-indentation instrument, i.e., depth-sensing indentation at elevated temperature, offers the capability to preciously control and monitor the load and penetration depth during the indentation process at the desired temperatures [10-12]. It is expected that the micro-indentation test can act as a convenient way to simulate micro embossing process and to study the mechanical response of the embossing materials at process temperature and applied pressure.

In this paper, micro-indentation tests using a micro flat-end cone indenter at various temperatures have been performed on green ceramic tapes to simulate imprinting process and investigate the mechanical response during imprinting. The correlation of indentation depth versus applied force at different temperatures was studied. The temperature and load-dependent creep and creep strain rate during holding period were obtained and discussed. The instantaneous recovery and retarded recovery after indentation were investigated as well.

## II. Experimental Methodology

### *A. Substrate materials*

Substrates laminated from low temperature co-fired ceramic (LTCC) green tapes were used in this study. The ceramic green tapes were HL2000 supplied by Heraeus. The ingredients of this ceramic green tape comprise ceramic-glass based powders, pores and polymeric additives. The thermogravimetric analysis (TGA) revealed that the amount of polymeric additives was approximately 16 wt%; the porosity factor was affected by the lamination parameters as well as casting process of the green tape. The thickness of a single layer green tape was 127 µm, and six layers of such green tapes were laminated under the pressure of 100 $kg/cm^2$ at the temperature of 70 °C for 10 minutes. Subsequently the laminated substrate with the total thickness being about 0.7 mm were cut into pieces of 4 mm x 7 mm for micro-indentation test.

### *B. Micro-indentation*

The mechanical test was performed with a micro-indentation system equipped with a high-temperature stage [12]. The sample stage itself included a heater, a thermocouple and ceramic block. A micro flat-end cone indenter was used in the experiment as shown in Fig. 1. A small heater and a miniature thermocouple were installed around the diamond indenter stub to prevent heat flow upon diamond-specimen contact. A thermal shield was interposed between the loading head and the hot stage to eliminate any thermal interference. The applied force was monitored by electric current applied to a movable coil, and the displacement of the indenter head was measured via a capacitive transducer. A small amount of high-temperature adhesive was used to attach specimen onto the hot stage. Some strips of glass fiber were used to isolate the diamond/specimen with environment to preclude heat flow between diamond/specimen and environment. The stage and the diamond indenter were heated up to the desired temperature simultaneously before indentation and the distance between indenter and specimen was 150 µm during the heating process. The loading duration was controlled at 20-30 seconds for all indentations and the holding time at the peak load was 300 seconds for investigating the creep response at various loads and temperatures. The test temperature was set to 25 °C and 75 °C, respectively. In each experiment, a holding period of 100 seconds was introduced after the load was unloaded to 10% of its peak value for the purpose of measuring instantaneous recovery and thermal drift at this unloading point.

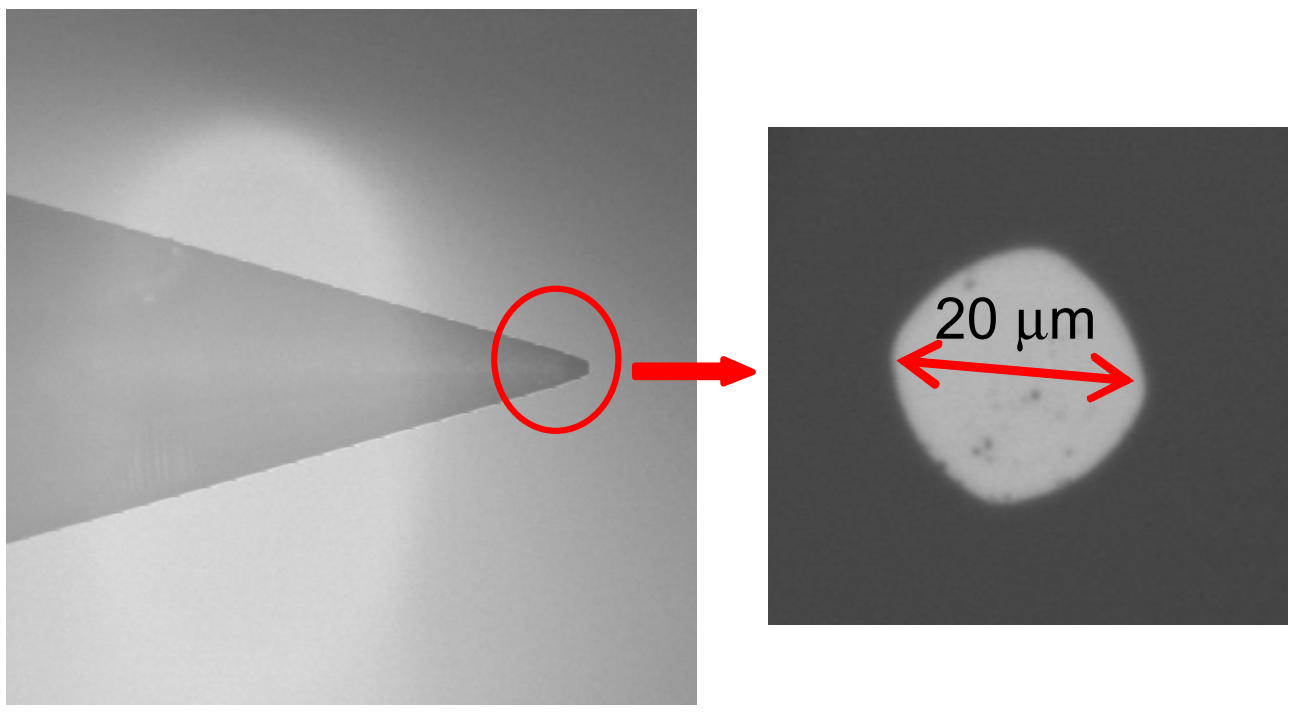

Fig. 1. The flat end cone indenter used in this study with a diameter of 20 µm and an angle of 39°48'11".

### *C. Charaterization*

After indentation, a Veeco atomic force microscope (AFM) was utilized to measure the profile of some shallower indentation cavities for studying the retarded recovery after indentation. The tapping mode was used with a vertical resolution of 0.1 nm. The scan area was set to 70 µm x 70 µm to cover an indentation cavity and the scan rate was 1 Hz.

## III. Results And Discussions

### *A. Indentation at various temperatures*

Fig. 2 shows the loading-hold-unloading curves of green ceramic tape at various peak loads and various temperatures. For example, curve ABCDEF represents the loading-hold-unloading curve at 75 °C under peak load of 8 mN. The portion of AB represents the loading process, the portion of BC represents the holding period at the peak load for 300 seconds, the portion of CDEF represents the unloading process in which the portion of DE represents the holding period of 100 seconds for measurement of the instantaneous recovery after 90% unloading and the thermal drift of the instrument. It was noticed that higher peak force caused deeper depth of penetration at a desired temperature, i.e., created a bigger cavity. It was also observed that similar depth of indentation could be generated with smaller force when working temperature was increased or dwell time at the peak load was longer. It indicated that the depth of indentation, i.e., size of cavity created, was force, temperature and dwell time dependent. It suggested that we could apply higher force at low temperature or apply lower force at higher temperature or apply smaller force at lower temperature but hold the peak load for longer time to create a particular size of cavity or channel. The cavity or channel could be retained after withdrawing the applied load. This was attributed to the plastic deformation of the green tape caused by compression. It implied that permanent feature could be formed by compression at low temperature even at room temperature.

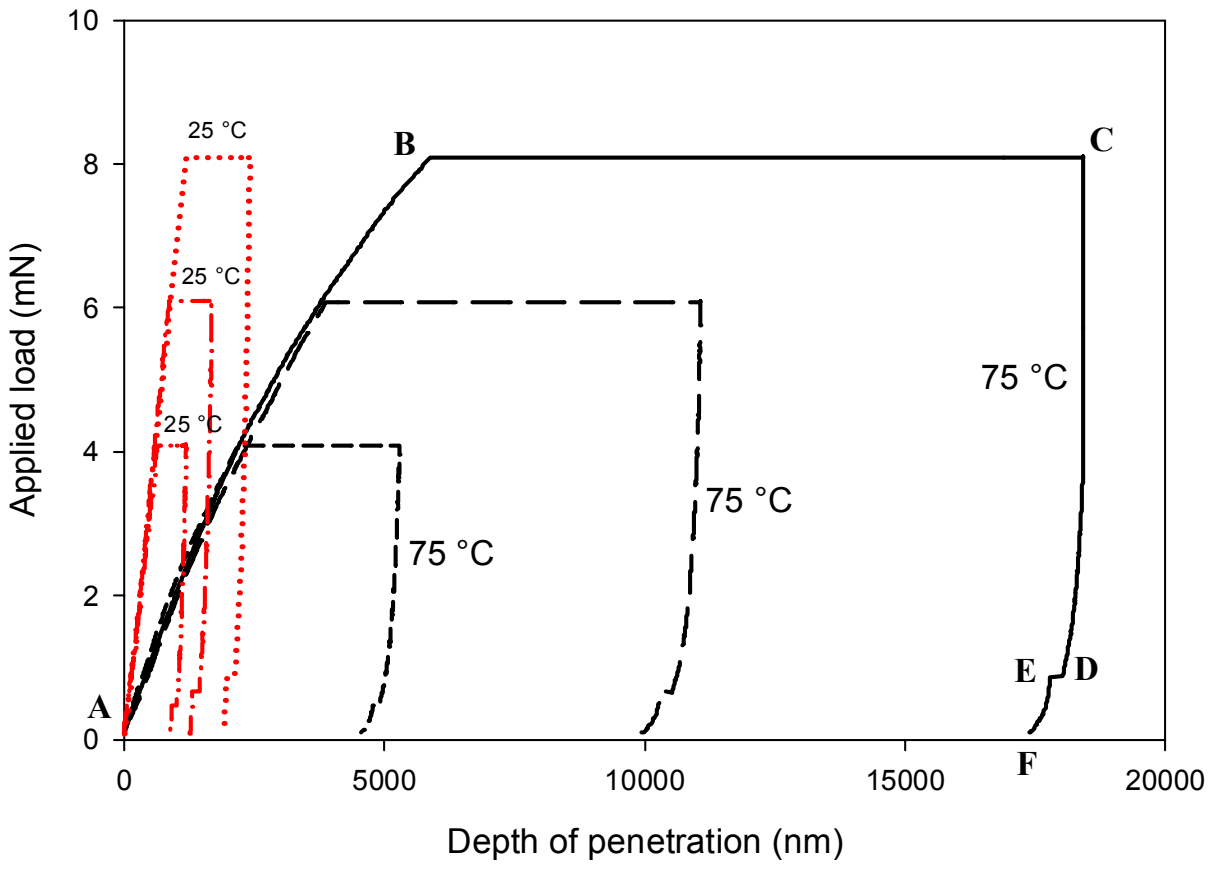

Fig. 2. The loading-hold-unloading curves of green substrate under different loads at 25 °C and 75 °C.

Figs. 3 and 4 show the applied force versus depth of penetration and the applied stress versus depth curves at 25 °C and 75 °C. There are three curves obtained from three locations for each temperature. The applied stress was calculated based on the following equation:

$$\sigma = \frac{P}{A} \quad (1)$$

where $\sigma$ is the stress, $P$ the force and $A$ is the projected contact area. $A$ is a function of indentation depth, $h$, and is given below in the present work.

$$A = \pi(10 + 0.362h)^2 \quad (2)$$

where the unit of $h$ is micrometer.

It was observed that the applied force and stress required for a particular depth of indentation at 75 °C were much smaller than those at 25 °C. It was also found that the indentation resistance was larger at the initial indentation, i.e., the incease of applied force and stress needed for penetrating a unit depth at the initial stage was larger than that at a deeper stage at both 25 °C and 75 °C.

*B. Creeps at various loads and temperatures*

From Fig. 2, it was noticed that there were significant increase in depth at the holding period under peak load. This implied that the creep occurred in the green tape. Fig. 5 shows the depth increase versus time under various loads and at various temepratures. The increase in depth at the peak load is defined as the creep penetration at the peak load. For micro-indentation, the creep strain rate can be expressed as the equation below [13]:

$$\frac{d\varepsilon}{dt} = \frac{1}{h}\frac{dh}{dt} \quad (3)$$

where $\frac{d\varepsilon}{dt}$ is the creep strain rate, $h$ the depth of indentation and $t$ is the creep time.

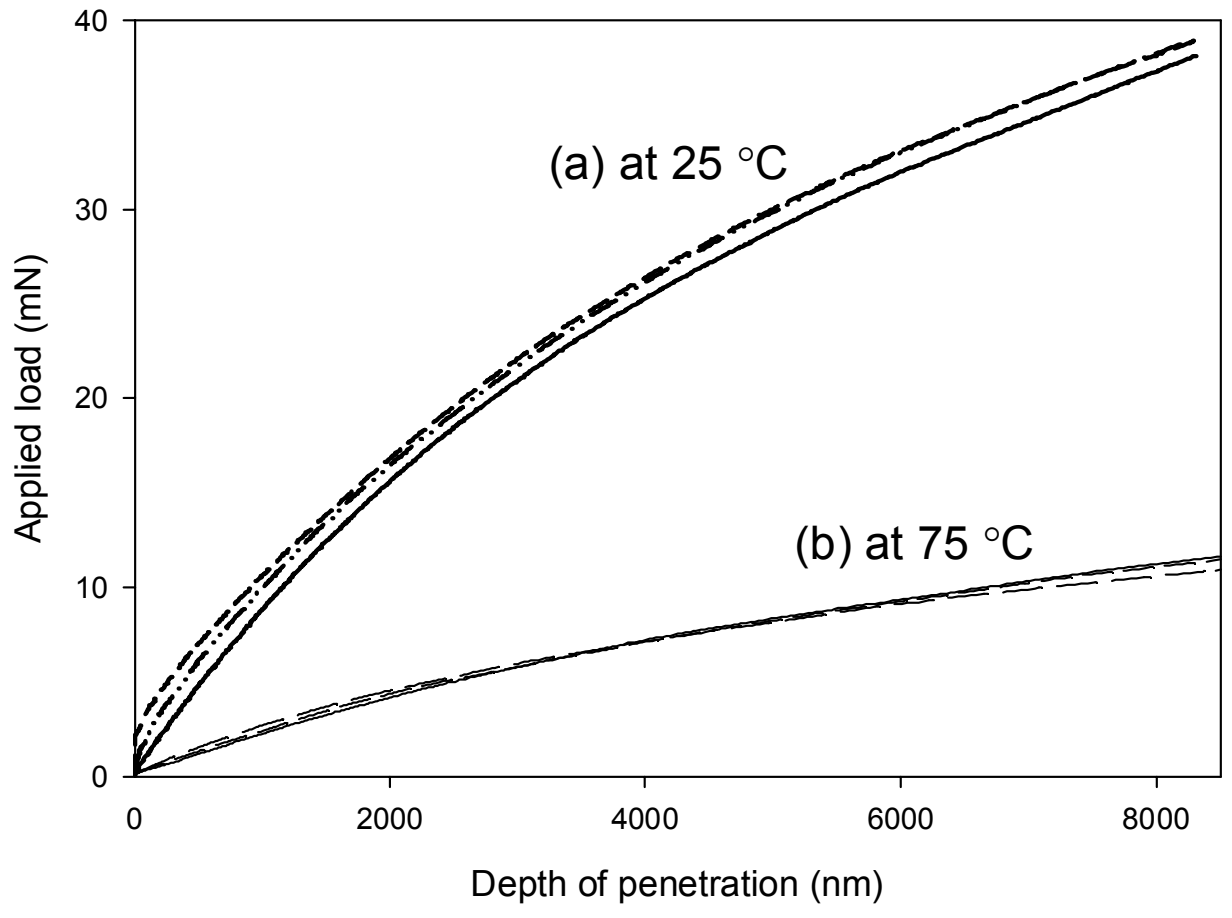

Fig. 3. Curves of applied force versus depth of penetration on ceramic green substrates at various temperatures: (a) at 25 °C; (b) at 75 °C.

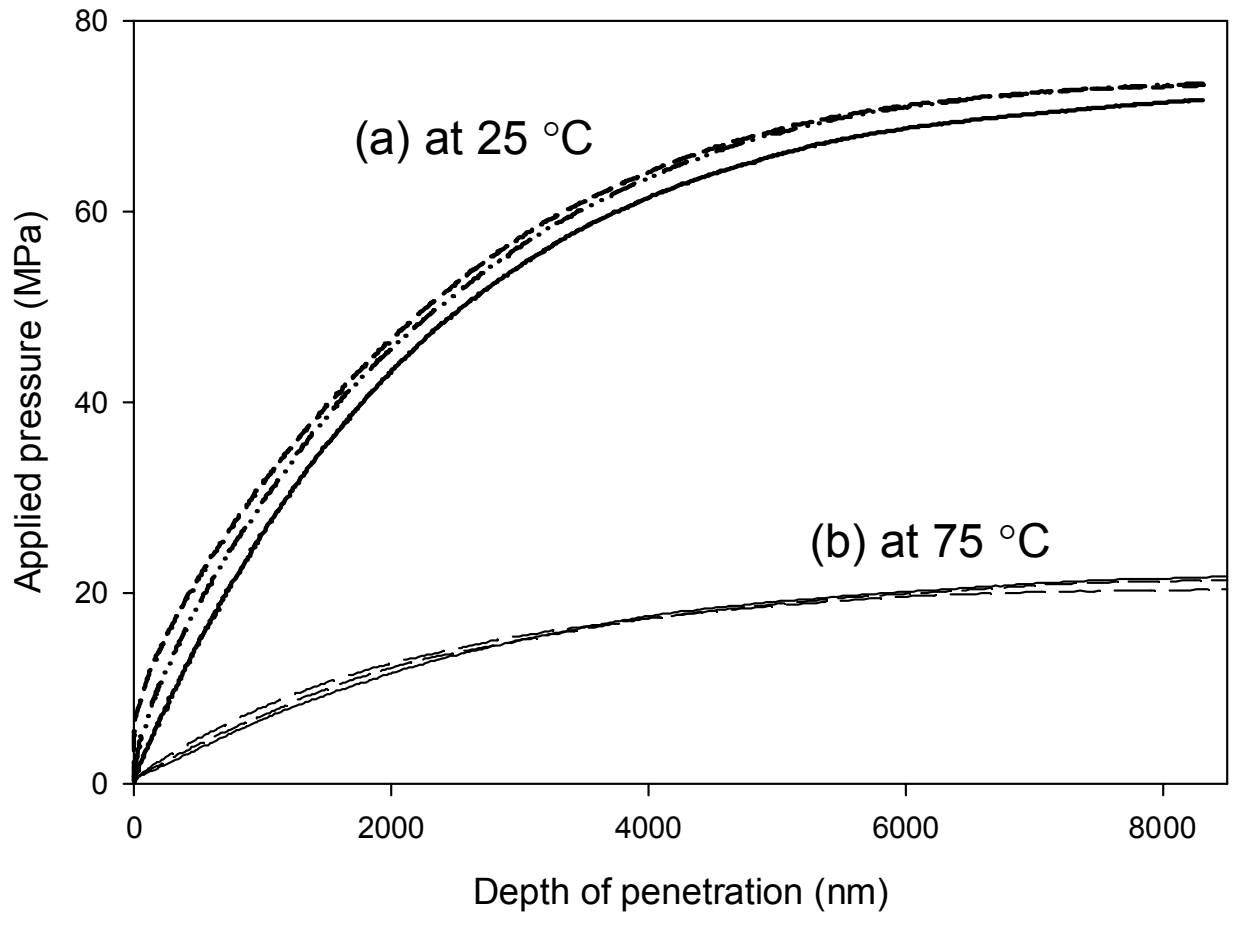

Fig. 4. Curves of applied pressure versus depth of penetration on ceramic green substrates at various temperatures: (a) at 25 °C; (b) at 75 °C.

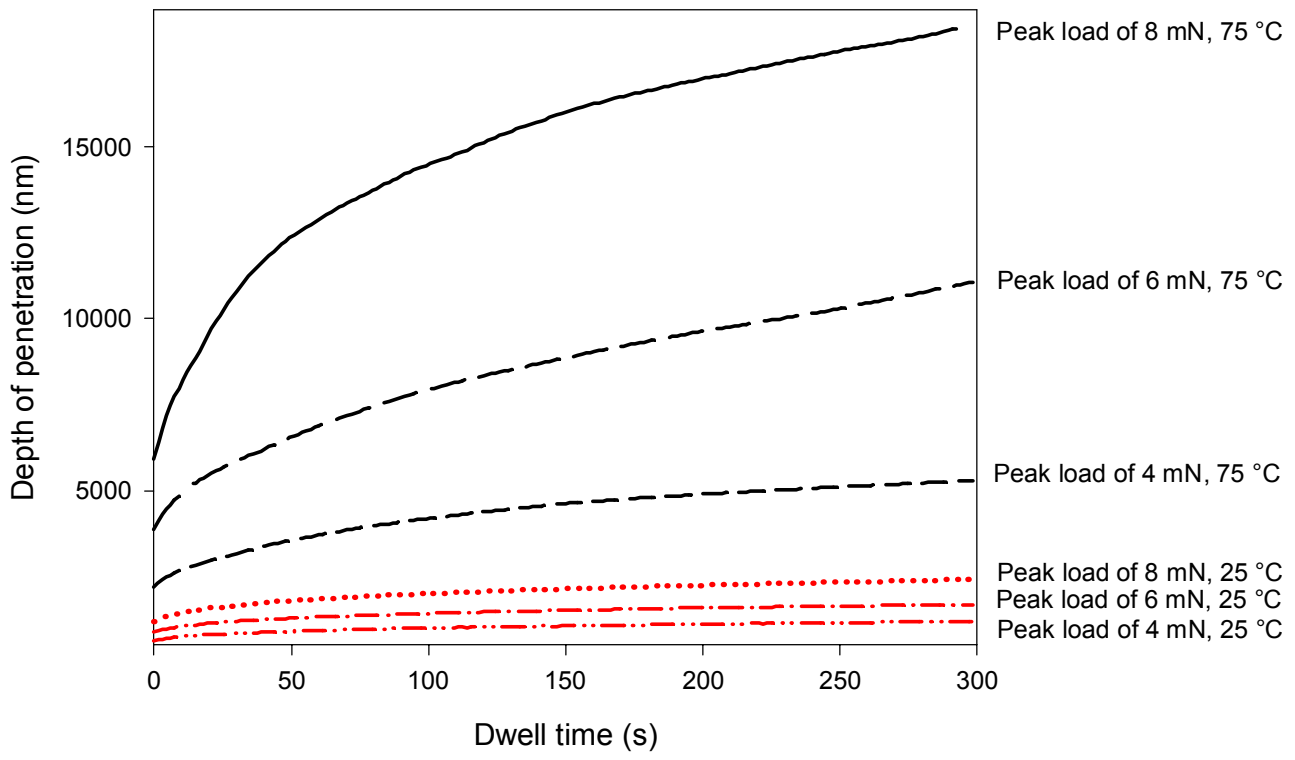

Fig. 5. Change of depth of penetration with the dwell time under various peak loads at 25 °C and 75 °C.

Based on the curves of creep penetration depth versus time, as shown in Fig. 5, the creep strain rate can be determined. Figs. 6 and 7 indicate the creep strain rate of the ceramic green substrate and all the curves are obtained from the curve fitting of the experimental results. It was noticed that the creep of green substrate was load and temeprtaure dependent. Higher load and higher temperature caused larger creep. It was also found that the creep strain rate at the peak load of 8 mN was the largest for both 25 °C and 75 °C. This suggested that larger peak load would cause larger creep strain rate. In addition, it was observed that green ceramic substrate exhibited higher creep strain rate at higher temperature; the creep strain rate for all temperatues and peak loads had a maximum value at the initial creep and it decreased dramatically and finally approached to zero after a period of load holding.

*C. Instantaneous recovery and retarded recovery when unloading*

In all unloading processes, we hold 10% of peak load for 100 seconds to monitor the change in depth after 90% unloading. It was observed that there was significant depth decrease at the holding period. For example, the depth of indentation decreased from 18017 nm (point D) to 17766 nm (point E) for the curve ABCDEF in Fig. 2. Fig. 8 shows the depth decrease versus time from point D to point E. It was noticed that the depth decreased significantly in the first 40 seconds and depth fluctuated from 40 seconds to 100 seconds. It could be interpreted as that the depth decrease in the first 40 seconds was due to the instantaneous recovery and the fluctuation of the depth after 40 seconds was caused by the thermal drift of the instrument. When the applied force was fully withdrawn, the depth of penetration approached to point F that is 17383 nm.

After indentation, an atomic force microscope was used to measure the profile of indentation cavities for studying the geometry of the indents and retarded recovery after indentation. The profile measurements were conducted both immediately after the specimen was removed from indentation instrument and a few days after indentation.

Fig. 9 shows the three-dimensional images of two indents measured as soon as specimens were removed from the indenter. Fig. 10 shows their cross-sectional profiles. It was observed that less load was required to obtain a deeper cavity at elevated temperature. It was also noticed that there was no obvious pile-up due to material flow near the edge regions of the indented cavities, which were normally observed in polymeric hot embossing. This indicated that, in the process of embossing ceramic green substrate, there was no significate material flow, or the material flow contributed to filling the neighboring porosities. This indicated that plastic deformation and viscoelastic deformation under compression were the main causes of generating micro features on ceramic green material at the temperatures of 25°C and 75 °C.

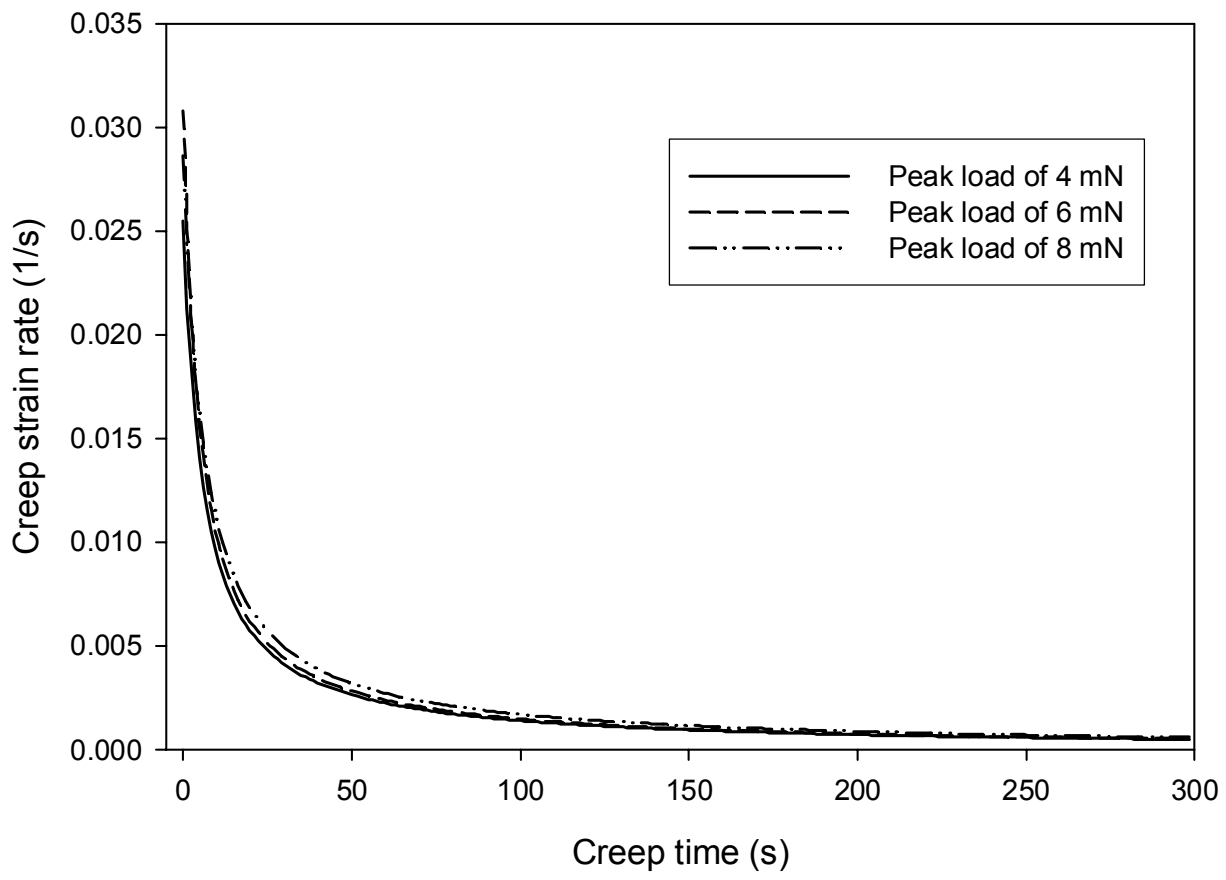

Fig. 6. Creep strain rate of the ceramic green substrate under various loads at 25 °C.

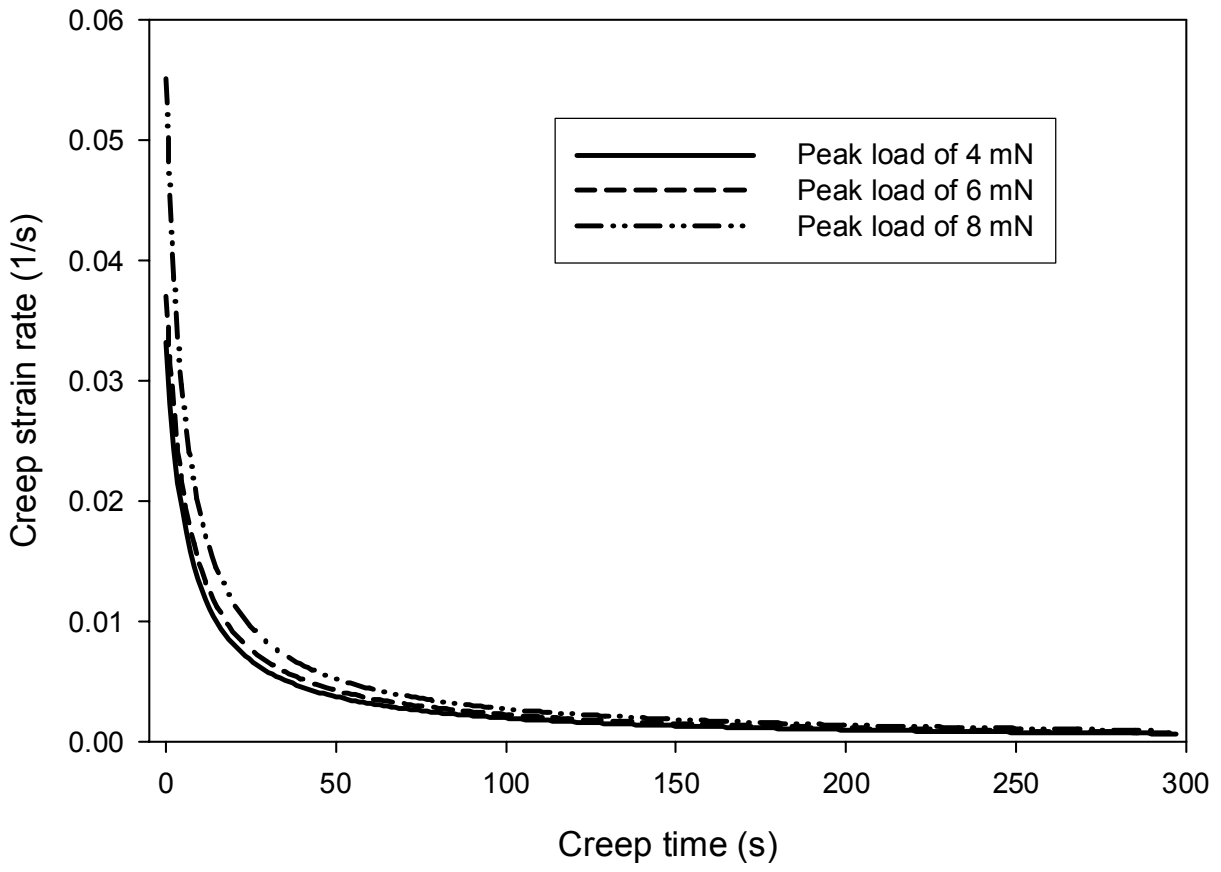

Fig. 7. Creep strain rate of the ceramic green substrate under various loads when heated up to 75 °C.

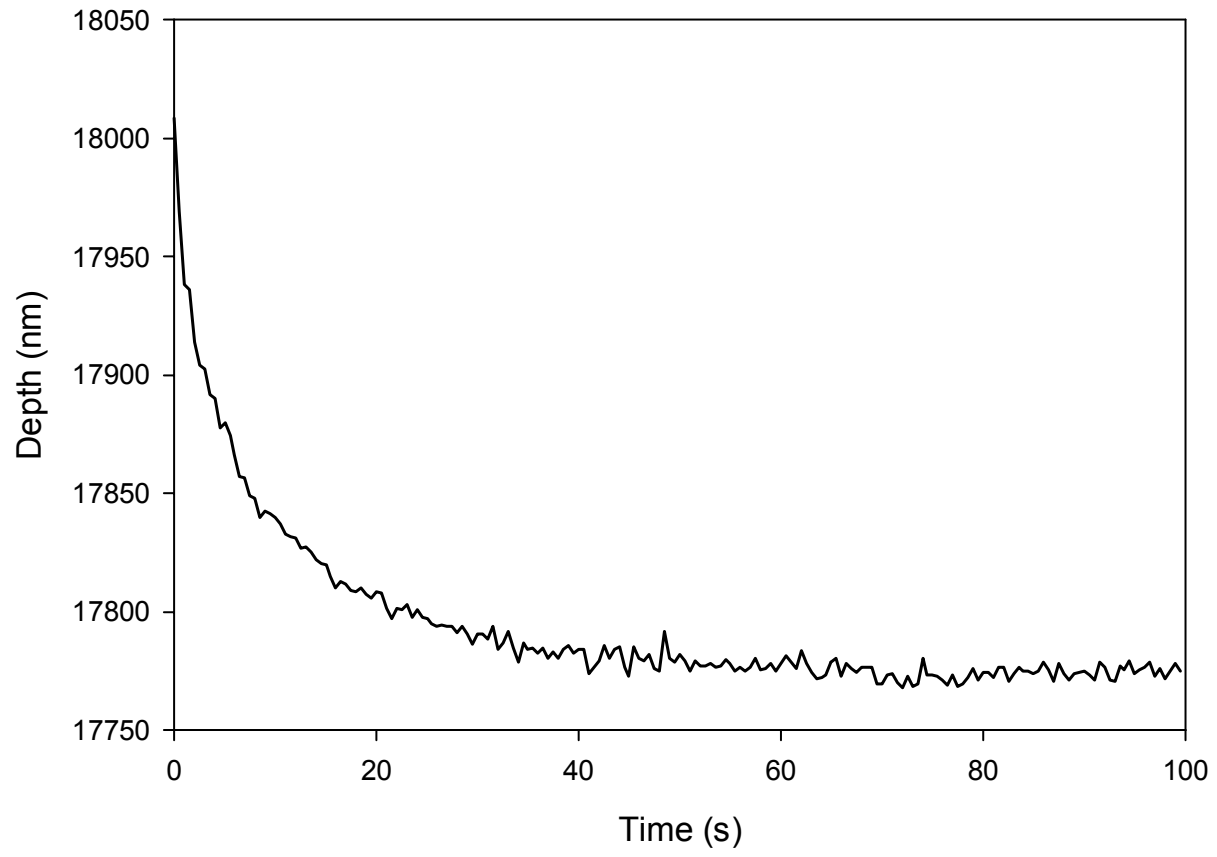

Fig. 8. The curve of indentation depth versus hold time as 90% of peak load (i.e. 8 mN) was unloaded at 75 °C. The significant depth decrease in the first 40 s was attributed to instantaneous recovery after 90% unloading and the fluctuation of the depth from 40 s to 100 s was due to thermal drift in the instrument.

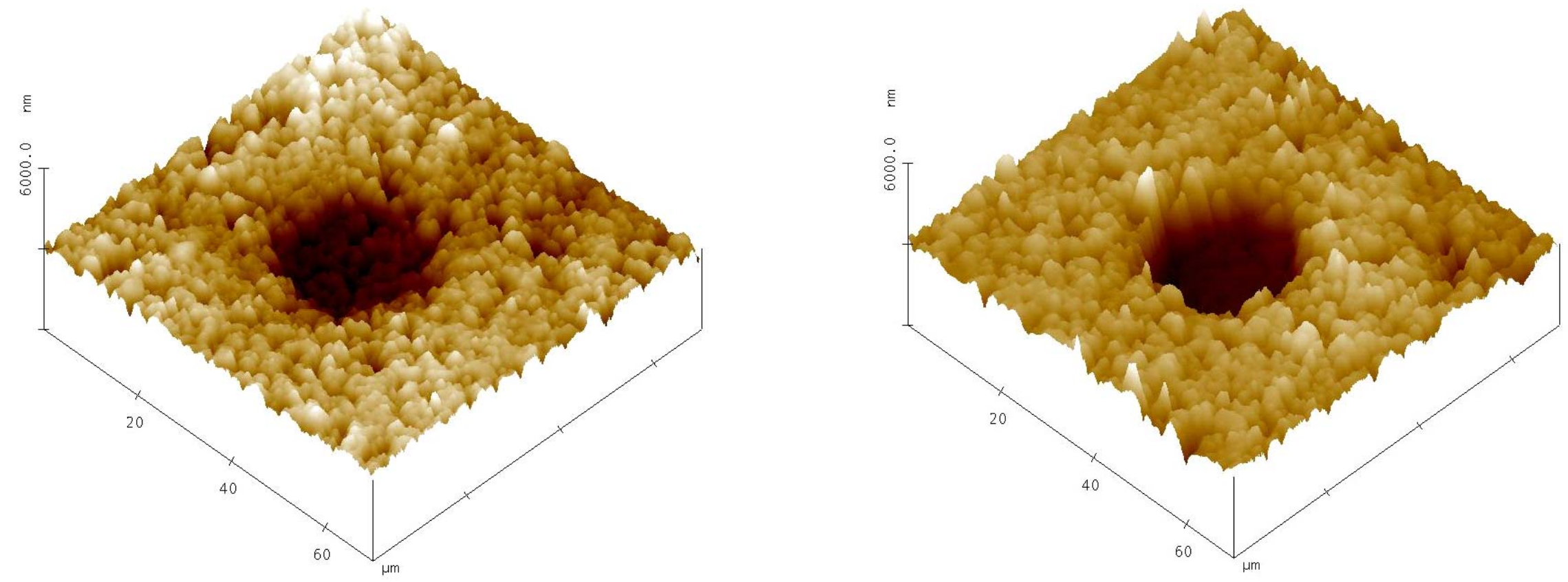

(a) Indent obtained at 25 °C and 10 mN (b) Indent obtained at 75 °C and 3.5 mN

Fig. 9. AFM images of indents obtained at various temperatures and applied loads.

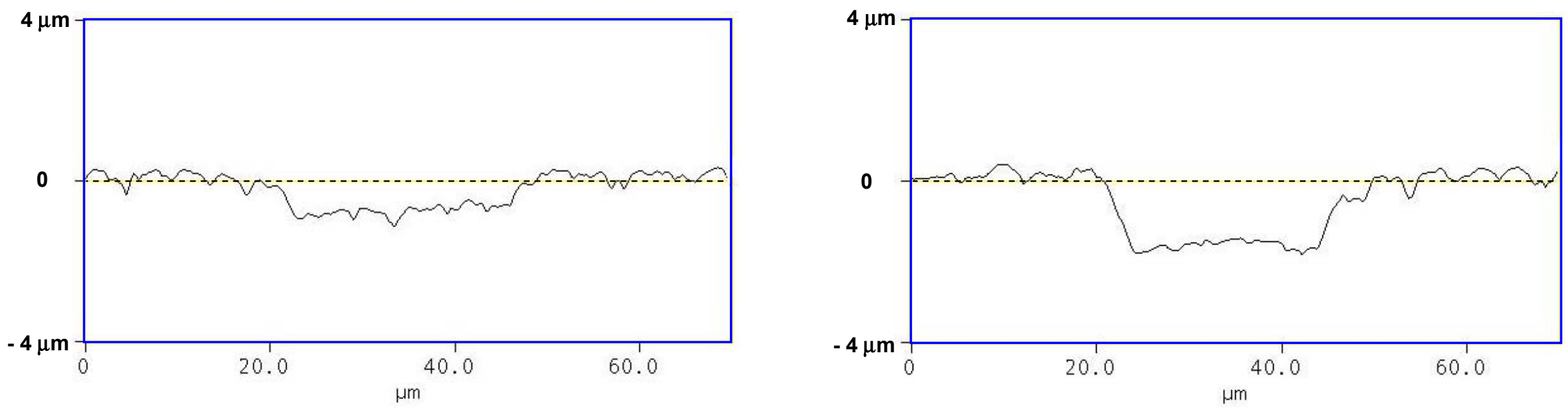

(a) An indented depth of 1 µm obtained at 25 °C and 10 mN (b) An indented depth of 2 µm obtained at 75 °C and 3.5 mN

Fig. 10. The cross-sectional profiles of indents obtained at different temperatures and applied loads

TABLE I
DEPTH CHANGE OF THE INDENTATION CAVITIES VERSUS TIME

| | Depth (µm) as -prepared | Depth (µm) after indentation for 1 day | Depth (µm) after indentation for 5 days |
|---|---|---|---|
| Cavity 1 | 2.29 | 2.14 | 2.13 |
| Cavity 2 | 2.27 | 2.17 | 2.18 |
| Cavity 3 | 2.26 | 2.15 | 2.07 |
| Ave. | 2.27 | 2.15 | 2.13 |

The reason for the little material flow in embossing could be the low amount of polymeric additives in ceramic green tapes. Table 1 indicates the depth change of the cavities after a period of indentation. It was found that the depth of the cavities reduced about 5% in the first day after indentation and the subsequent decrease was less than 1% which is considered as insignificant. It can be concluded that the recovery occured immediate after indentation and there was no significant retarded recovery in the depth of cavity after the first day of indentation.

## IV. CONCLUSIONS

Thermo-mechanical behavior of ceramic green substrate has been studied by micro-indentation experiments at 25 °C and 75 °C. Indentation cavity could be formed in the temperature range from 25 °C to 75 °C and the cavity formation was attributed to the plastic deformation and viscoelastic deformation of green ceramic substrate under compression. The size of the cavities created was force, temperature and dwell time dependent. Higher temperature, higher applied force and longer dwell time could generate bigger and deeper indentation cavities. Creep occurred and made a larger contribution to the plastic deformation during the indentation process at elevated temperatures and high peak loads. There was instantaneous recovery during the unloading process and retarded recovery in the first day after indentation. The permanent cavity could be retained after indentation. The results could be used as a guideline for embossing ceramic green substrates.

ACKNOWLEDGMENT

The authors would like to thank Ms. Hla Phone Maw for sample preparation. This work is supported by Agency for Science, Technology and Research (A*STAR) of Singapore for Singapore-Poland Cooperation.